# Technically natural Higgs boson from Planck scale

Martin Rosenlyst[*]
*Rudolf Peierls Centre for Theoretical Physics, University of Oxford,
1 Keble Road, Oxford OX1 3NP, United Kingdom
and CP³-Origins, University of Southern Denmark,
Campusvej 55, DK-5230 Odense M, Denmark*

We propose UV complete (partially) composite Higgs models with compositeness scale up to the Planck scale assisted by a novel mechanism. This mechanism is based on softly breaking a global $\mathbb{Z}_2$ symmetry by technically natural small vacuum misalignment, dynamically triggering the electroweak symmetry breaking and Standard Model fermion mass generation. This mechanism can be present in various models based on vacuum misalignment. For concreteness, we demonstrate it in a minimal partially composite two-Higgs boson scheme, where the Higgs boson is a mixture of a composite and an elementary state, transforming odd under a $\mathbb{Z}_2$ symmetry. For this concrete model example, all dimensionful fundamental parameters are approximately $\mathcal{O}(10^{18})$ GeV. We study the vacuum stability of this model by investigating the renormalization group running of the quartic coupling of the Higgs boson. Furthermore, the parameter space can already be searched by gravitational waves from a confinement-induced phase transition. Finally, the mass and mixing of the neutrinos may be naturally generated via loops of a second Higgs doublet transforming even under the $\mathbb{Z}_2$ symmetry, which may be challenged by lattice calculations and a more accurate measurement of the top mass.

## I. INTRODUCTION

Fundamental strongly coupled gauge Yukawa models with a strongly interacting fermion sector were proposed in Ref. [1] for electroweak symmetry breaking (EWSB) and Standard Model (SM) fermion mass generation. The motivation was to achieve dynamical EWSB and to alleviate the SM naturalness problem. A challenging aspect in this framework is, however, the fermion mass generation, where various ideas have been proposed to address this, for example, with the following approaches:

  (i) extended technicolor (ETC) [2],
  (ii) partial compositeness (PC) [3],
  (iii) fundamental partial compositeness (FPC) [4,5],
  (iv) partially composite Higgs (PCH) [6,7].

However, these fermion mass generation approaches may suffer from problems due to, for example, reintroduction of new naturalness problems, generation of dangerous flavor changing neutral currents (FCNCs), or instability of the Higgs vacuum. In the following, we will discuss these types of issues for the different approaches before we present a novel mechanism alleviating these problems.

For the two former fermion mass generation approaches (the ETC and PC approaches), the SM fermions couple to the strong sector arising via higher-dimension operators: for example, bilinear ETC-type and linear PC operators of the SM fermions of the form $ff\Psi\Psi$ and $f\Psi\Psi\Psi$, respectively. Here the fields $f$ and $\Psi$ represent, respectively, one of the SM fermions and the new strongly interacting fermions (hyperfermions). Those effective couplings may arise from the exchange of heavy scalar multiplets or heavy vectors at different "flavor" scales $\Lambda_f$ corresponding to the SM fermions $f$. For the ETC approach [2], those operators are responsible for generating the SM-fermion masses $m_f$ and Higgs-Yukawa couplings $y_f$ typically as follows [8]:

$$m_f = y_f \frac{v_{\text{EW}}}{\sqrt{2}} \sim g_{f_L} g_{f_R} \left(\frac{\Lambda_{\text{HC}}}{\Lambda_f}\right)^2 v_{\text{EW}}, \quad (1)$$

where $v_{\text{EW}} \approx 246$ GeV is the EW vacuum expectation value (VEV), $g_{f_{L/R}}$ are the couplings between the left- and right-handed SM fermions, respectively, and their corresponding heavy scalars or vectors, and $\Lambda_{\text{HC}}$ is the compositeness scale of the composite dynamics by a new strong hypercolor (HC) gauge group $G_{\text{HC}}$. However, the main problem in the ETC approach is the generation of dangerous FCNCs, since the flavor scale for the top quark must be small enough to generate the top mass. However, in the old ETC models, it was assumed that the large top mass would be generated by the presence of an infrared conformal phase [9] relying on the property that the strong sector enters a "walking" phase [10] above the compositeness scale. Assuming a new strong coupling conformal from the flavor scale of the top quark $\Lambda_t$, down to $\Lambda_{HC}$, the top mass renormalized at $\Lambda_t$ is modified by [8]

$$m_t(\Lambda_t) \simeq m_t(\Lambda_{HC})\left(\frac{\Lambda_t}{\Lambda_{HC}}\right)^{-\gamma_m}, \qquad (2)$$

where $\gamma_m$ is the anomalous mass dimension of the fermion bilinear, which is nonperturbatively determined. From the unitarity bounds of scaling dimensions of these operators, we have $\gamma_m \geq -2$ [11]. The walking dynamics can thus lift the fermion mass and suppress the FCNCs by increasing $\Lambda_t$ due to the fact that the FCNC terms are suppressed by $\sim \Lambda_t^{-2}$ [8]. However, according to Ref. [12], the anomalous mass dimension can be a maximum of $\gamma_m \approx -2$ to achieve the observed top mass, which is very close to the minimum value regarding the unitarity bound. It is, therefore, difficult to obtain the observed top mass with ETC-type operators without generating a severe amount of FCNCs.

---

[*]martin.jorgensen@physics.ox.ac.uk

An alternative fermion mass generation approach revived in the holographic model is the PC approach [3], where the top quark features a linear coupling to the strong sector of the form $f\Psi\Psi\Psi$. This approach helps to avoid FCNCs and generate a large enough top mass due to the relaxed unitarity bounds of the anomalous dimensions of operators consisting of three hyperfermions, $\gamma_m \geq -3$. Therefore, we need $-2 \gtrsim \gamma_m \gtrsim -3$ [12] in this scenario to generate the observed top mass. This enhancement from large anomalous dimensions of the fermionic operators allows us to push the flavor scale high enough without suppressing the SM-fermion mass operators. The first realistic attempt that provides a UV completion of CH models was proposed in Refs. [13,14], where the SM-fermion masses are generated based on a PC mechanism. In this work, the Techni-Pati-Salam CH model is presented, based on a renormalizable gauge theory $SU(8)_{PS} \otimes SU(2)_L \otimes SU(2)_R$. This gauge group spontaneously breaks in several steps down to $G_{HC} \otimes G_{SM}$ with the new strong gauge group $G_{HC} = Sp(4)_{HC}$ and the SM gauge group $G_{SM}$, resulting in different heavy gauge bosons and scalars. However, the realization of this framework turns out to be highly nontrivial in practice due to the theoretical and phenomenological requirements. In the following, the main motivation will, therefore, be to construct a more simple UV completion, which is easier to realize. Before considering such model examples, we consider the latter two fermion mass generation approaches that do not include effective operators, namely, the FPC and PCH approaches.

For the FPC approach [4,5], the fermion masses are generated via fundamental Yukawa couplings of the form $f\Psi S$, involving scalars $S$ charged under the new strong gauge interactions, $G_{HC}$. Unfortunately, the FPC models are not free from hierarchy problems, since the new strongly interacting scalars (hyperscalars) need to achieve their mass from new Higgs-like mechanisms. Regarding Ref. [5], a fermion mass generated by this approach is roughly given by $m_f \sim (\Lambda_{HC}/M_{S_f})^2 v_{EW}$, where $M_S$ is the mass of the hyperscalar $S_f$ corresponding to the SM-fermion mass $m_f$. Therefore, the top mass arises from the Yukawa coupling with the lightest hyperscalar, requiring a mass which is only a few orders of magnitude larger than the compositeness scale. Thus, these models can maximally alleviate the SM naturalness problem by a few orders of magnitude. However, this naturalness problem can be further alleviated by pushing the compositeness scale beyond the TeV scope, allowing larger $M_{S_f}$ without suppressing the top mass, but this procedure induces a new unnatural hierarchy between the compositeness and EW scale. In this paper, we thus suggest a novel mechanism where this new hierarchy can be technically natural explained by softly breaking a global $\mathbb{Z}_2$ symmetry of the new strong interactions. Moreover, implementation of this mechanism in models with ETC-type and PC operators, even without a walking phase, can enable a much larger $\Lambda_{HC}$ in Eq. (1) in a technically natural way, resulting in suppression of the FCNCs and easier construction of a UV complete theory.

Finally, we consider the PCH approach [6,7,15] generating the fermion masses in terms of fundamental Yukawa interactions of the forms $\Psi\Psi\Phi$ and $\bar{f}fH$ with $H \in \Phi$, where $\Phi$ is an "elementary" scalar multiplet. In this framework, the Higgs doublet is a mixture of a composite doublet and the elementary $SU(2)_L$ doublet $H$. Via the Yukawa interactions of the form $\Psi\Psi\Phi$, the VEV from the vacuum misalignment in the composite sector can be transferred to the neutral $CP$-even component of the elementary doublet $H$, providing fermion masses via SM-like Yukawa couplings of the form $\bar{f}fH$. However, the renormalization group (RG) running analysis performed in Ref. [16] shows that these models suffer from a low vacuum instability scale. Supposing stability of the vacuum in these models up to a minimum of the compositeness scale, the mass parameter of the scalar multiplet $\Phi$ can only be pushed to values of a few orders of magnitude larger than the SM one. Therefore, these models can only provide a weak alleviation of the SM naturalness problem. Similarly, for this model type, the SM naturalness problem can be alleviated by a large compositeness scale based on the novel mechanism presented here.

Meanwhile, we have learned that a large compositeness scale can alleviate the issues of the various fermion mass approaches, which can be technically natural due to a novel mechanism. Realization of this novel mechanism can be achieved in models based on vacuum misalignment with an associated $\mathbb{Z}_2$ symmetry, providing a minimum of one

Higgs doublet that is odd under the $\mathbb{Z}_2$ symmetry. By vacuum aligning in the direction of the neutral CP-even component of this doublet with a small angle $\theta$ resulting in a large compositeness scale, the global $\mathbb{Z}_2$ symmetry is very weakly broken. According to 't Hooft's naturalness principle [1], this small angle is technically natural due to the fact that the $\mathbb{Z}_2$ symmetry is restored for a vanishing angle. Although this is technically natural, then it is important to note that no dynamical rationale for the smallness of the vacuum misalignment angle $\theta$ is offered. In this paper, the Higgs boson will arise as a pseudo-Nambu-Goldstone boson (PNGB) from a spontaneously broken global symmetry. The geneology of Higgs bosons as PNGBs includes "composite Higgs" (CH) bosons [17], PCH bosons [6,7,15,16], "little Higgs" bosons [18,19], "holographic extra dimensions" [20,21], and "twin Higgs" bosons [22]. This dynamics may also be realized in "elementary Goldstone Higgs models" [23] containing only elementary scalars. In such a model, we need to add at least 26 real scalars[1] beyond the SM compared to only six Weyl fermions strongly interacting under a new gauge group in the following model example. Furthermore, there are fewer parameters in the model example presented here. We leave the study of this class of models to future work. For concreteness, throughout this work, we consider this mechanism in the (P)CH framework, where the small vacuum misalignment angle results in a compositeness scale up to the Planck scale, leading to potentially a complete alleviation of the SM naturalness problem, suppression of FCNCs, and stabilization of the vacuum.

The paper is organized as follows: In Sec. II, we show the road to UV complete (P)CH models assisted by the novel mechanism. In Sec. III, we demonstrate this mechanism in a concrete UV complete PCH model, predicting the Higgs boson as a partially composite particle. In Sec. IV, we study the vacuum stability of this concrete model. In Sec. V, we search the parameter space of this model by gravitational waves from a confinement-induced phase transition. In Sec. VI, we investigate how neutrino masses may be loop induced in this model. Finally, in Sec. VIII, we give our conclusions.

## II. THE ROAD TO UV COMPLETE (P)CH MODELS

We presently focus on CH models with vacuum misalignment based on an underlying gauge description of hyperfermions. Different chiral symmetry breaking patterns in these CH models are discussed in Refs. [24,25], and we note the following minimal cosets with a Higgs candidate and custodial symmetry: SU(4)/Sp(4) [26], SU(5)/SO(5) [27], SU(6)/Sp(6) [28], SU(6)/SO(6) [29], and SU(4) × SU(4)/SU(4) [30]. A $\mathbb{Z}_2$ symmetry is present in the three latter cases [28,29,31]. For concreteness, we consider, therefore, the minimal composite Higgs model with one $\mathbb{Z}_2$-odd composite Higgs doublet, which is the model example with the coset SU(6)/Sp(6) [28] consisting of two SU(2)$_L$ doublets and two singlets of Weyl hyperfermions. With this setup, it is technically natural to have a small misalignment in the direction of the neutral CP-even component of the $\mathbb{Z}_2$-odd composite Higgs doublet by balancing precisely the contributions to the Higgs potential from the top loops, gauge loops, and the explicit masses of the hyperfermions. By assuming that the compositeness scale is set to the Planck scale $\Lambda_{HC} \sim 4\pi f = m_{Planck}$, the vacuum misalignment angle is $\theta \approx 2.5 \times 10^{-16}$ according to the expression of the EW VEV $v_{EW} = f s_\theta$, where $f$ is the GB decay constant of the composite sector. So far, this model example completely ameliorates the SM naturalness problem. However, we need to specify a UV complete theory responsible for generating the SM-fermion masses without introducing new problems.

In this paper, we employ the PCH fermion mass generation approach [6,7,15,16] to generate the SM-fermion masses. We leave the specifics of possible UV complete theories with the other fermion mass generation approaches for future work. Instead of adding effective operators with ETC and PC or several hypercolored scalars with FPC approaches, we consider the simplest UV completion of this SU(6)/Sp(6) CH model, where we keep the SU(2)$_L$ elementary Higgs doublet $H$ in the SM. Therefore, we only need to add six extra Weyl fermions strongly interacting under a new strong gauge group $G_{HC}$, which may be the minimal SU(2)$_{HC}$ gauge group. Different from the SM, we assume that $H$ is odd under the $\mathbb{Z}_2$ symmetry of the composite sector, leading to that the Higgs boson arises as a mixture between the $\mathbb{Z}_2$-odd composite PNGB from the spontaneously global symmetry breaking SU(6) → Sp(6) and the elementary weak doublet $H$. Furthermore, via new Yukawa interactions between the strongly interacting hyperfermions and $H$, the VEV generated by the vacuum misalignment in the composite sector can be transferred to the neutral CP-even component of $H$, leading to a VEV $v$ of it. Based on that, the SM fermions can achieve their masses via ordinary Yukawa couplings to $H$. On the other hand, the EW gauge bosons obtain masses from both the VEV of the elementary Higgs doublet and the vacuum misalignment in the composite sector such that the EW scale is set by

$$v_{EW}^2 = v^2 + f^2 \sin^2\theta, \quad (3)$$

where $\theta$ ($\pi/2 \leq \theta \leq \pi$) parametrizes the vacuum misalignment. At $\sin\theta = 0$ ($\theta = \pi$), the EW and $\mathbb{Z}_2$ symmetries are unbroken, while at $\sin\theta = 1$ ($\theta = \pi/2$) the condensate is purely SU(2)$_L$ and the $\mathbb{Z}_2$ symmetry is broken (this limit is commonly referred to as bosonic technicolor proposed and explored in Refs. [32–36]). This concrete PCH model example is thus technically natural even though the vacuum misalignment angle $\theta$ is very close to $\pi$ since the $\mathbb{Z}_2$ symmetry is restored for $\theta = \pi$. The following section considers this concrete UV complete model in detail.

---

[1]There exist 14 PNGBs in the SU(6)/Sp(6) coset ($\Pi_i$ with $i = 1, ..., 14$), which is the minimal coset where the mechanism presented here can be realized. Regarding the scalar matrix in Eq. (14) in Ref. [23], we need the scalars $\sigma$, $\Theta$, $\Pi_i$, and $\tilde{\Pi}_i$ (i.e., 30 scalars in total) to realize the mechanism in an elementary Goldstone Higgs model, where four of them will go to the Higgs doublet.

## III. A CONCRETE PARTIALLY COMPOSITE HIGGS MODEL

In the following, we focus on the concrete SU(6)/Sp(6) model [28] explored in Refs. [37–41] with one elementary $\mathbb{Z}_2$-odd Higgs doublet $H$ as a template for this mechanism. We assume four Weyl fermions are arranged in two SU(2)$_L$ doublets $\Psi_1 \equiv (\psi_1, \psi_2)^T$ and $\Psi_2 \equiv (\psi_5, \psi_6)^T$, and two in SU(2)$_L$ singlets $\psi_{3,4}$ with hypercharges $\mp 1/2$. We have listed in Table I the representations of the gauge groups and parity under the $\mathbb{Z}_2$ symmetry of the fermions and the elementary weak doublet in the model.

### A. The condensate and PNGBs

By assuming the Weyl hyperfermions are a fundamental representation of the new strongly interacting gauge group $G_{HC} = SU(2)_{HC}$ or $Sp(2N)_{HC}$, which is the pseudo-real representation, we can then construct an SU(6) flavor multiplet by arranging the six Weyl hyperfermions into an SU(6) vector $\Psi \equiv (\psi^1, \psi^2, \psi^3, \psi^4, \psi^5, \psi^6)^T$. This results in the chiral symmetry breaking SU(6) $\rightarrow$ Sp(6) when the hyperfermions confine. The hyperfermions develop a nontrivial and antisymmetric vacuum condensate [26]

$$\langle \Psi^I_{\alpha,a} \Psi^J_{\beta,b} \rangle \epsilon^{\alpha\beta} \epsilon^{ab} \sim E^{IJ}_{CH}, \quad (4)$$

where $\alpha, \beta$ are spinor indices, $a, b$ are HC indices, and $I, J$ are flavor indices. We will suppress the contractions of these indices for simplicity. The vacuum of the composite sector giving rise to the VEV of the neutral CP-even component $\Phi^0_{odd}$ of the $\mathbb{Z}_2$-odd composite doublet, by misalignment, can be written as [26]

$$E_{CH} = \begin{pmatrix} +i\sigma_2 & 0 & 0 \\ 0 & -i\sigma_2 c_\theta & -\mathbb{1}_2 s_\theta \\ 0 & +\mathbb{1}_2 s_\theta & +i\sigma_2 c_\theta \end{pmatrix}, \quad (5)$$

where $\sigma_2$ is the second Pauli matrix, and from now on we use the definitions $s_x \equiv \sin x$, $c_x \equiv \cos x$, and $t_x \equiv \tan x$.

The chiral symmetry breaking SU(6) $\rightarrow$ Sp(6) results in 14 PNGBs, $\pi_a$ with $a = 1, ..., 14$, and thus, 14 SU(6) broken generators $X_a$ correspond to the vacuum $E_{CH}$. The Goldstone bosons around the CH vacuum $E_{CH}$ are parametrized as

$$\Sigma(x) = \exp\left[\frac{2\sqrt{2}i}{f}\pi_a(x)X_a\right] E_{CH} \quad (6)$$

with the decay constant $f$ of them. This model preserves a $\mathbb{Z}_2$ symmetry generated by an SU(6) matrix, which is

$$P = \text{Diag}(1, 1, 1, 1, -1, -1), \quad (7)$$

where the $\mathbb{Z}_2$-odd fields of the model are the composite PNGBs:

$$\Phi^0_{odd}, \quad (\Phi^0_{odd})^*, \quad \Phi^\pm_{odd}, \quad \Delta^0, \quad \Delta^\pm, \quad \varphi^0. \quad (8)$$

We have listed in Table II the quantum numbers for the EW unbroken vacuum ($s_\theta = 0$) and $\mathbb{Z}_2$ parity of the PNGBs divided into the groupings: the $\mathbb{Z}_2$-even PNGBs in the minimal $G_0/H_0 = SU(4)/Sp(4)$ CH subset [26], and the additional $\mathbb{Z}_2$-odd and -even PNGBs in the rest of the SU(6)/Sp(6) subset. The neutral components of the composite weak doublets $\Phi_{even, odd}$ are as follows:

$$\Phi^0_{even} \equiv \frac{\phi_R - i\phi_I}{\sqrt{2}}, \quad \Phi^0_{odd} \equiv \frac{h - i\pi^3}{\sqrt{2}}, \quad (9)$$

while the elementary $\mathbb{Z}_2$-odd doublet is written as

$$H = \frac{1}{\sqrt{2}}\begin{pmatrix} \sigma_h - i\pi^3_h \\ -(\pi^2_h + i\pi^1_h) \end{pmatrix}. \quad (10)$$

### B. The chiral Lagrangian and the effective potential

With this field content and vacuum of this specific PCH model, the underlying Lagrangian describing the new

TABLE I. The hyperfermions in the SU(6)/Sp(6) template model and the elementary Higgs doublet $H$ labeled with their representations of $G_{HC} \times SU(3)_C \times SU(2)_L \times U(1)_Y$ and parity under the $\mathbb{Z}_2$ symmetry.

|  | $G_{HC}$ | SU(3)$_C$ | SU(2)$_L$ | U(1)$_Y$ | $\mathbb{Z}_2$ |
|---|---|---|---|---|---|
| $\Psi_1 \equiv (\psi_1, \psi_2)^T$ | □ | 1 | □ | 0 | $+1$ |
| $\psi_3$ | □ | 1 | 1 | $-1/2$ | $+1$ |
| $\psi_4$ | □ | 1 | 1 | $+1/2$ | $+1$ |
| $\Psi_2 \equiv (\psi_5, \psi_6)^T$ | □ | 1 | □ | 0 | $-1$ |
| $H$ | 1 | 1 | □ | $+1/2$ | $-1$ |

TABLE II. The PNGBs in the SU(6)/Sp(6) template model in the EW unbroken vacuum ($s_\theta = 0$) labeled with their $(SU(2))_L, U(1)_Y)_{\mathbb{Z}_2}$ quantum numbers.

| $G_0/H_0$ | $\mathbb{Z}_2$-odd PNGBs | $\mathbb{Z}_2$-even PNGBs |
|---|---|---|
| $\Phi_{even} = (2, 1/2)_+$ | $\Phi_{odd} = (2, 1/2)_-$ | $\eta = (1, 0)_+$ |
| $\eta' = (1, 0)_+$ | $\Delta = (3, 0)_-$ |  |
|  | $\varphi^0 = (1, 0)_-$ |  |

strong sector and the elementary doublet can be written as [16]

$$\mathcal{L}_{\text{PCH}} = \Psi^\dagger i\gamma^\mu D_\mu \Psi + D_\mu H^\dagger D^\mu H$$
$$- m_H^2 H^\dagger H - \lambda_H (H^\dagger H)^2 - \left(\frac{1}{2}\Psi^T M \Psi\right.$$
$$\left. + y_U H_\alpha (\Psi^T P^\alpha \Psi) + y_D \tilde{H}_\alpha (\Psi^T \tilde{P}^\alpha \Psi) + \text{H.c.}\right) \quad (11)$$

with $\tilde{H} \equiv \epsilon H^*$. We have introduced the spurions

$$2P_{ij}^1 = \delta_{i5}\delta_{j3} - \delta_{i3}\delta_{j5}, \quad 2P_{ij}^2 = \delta_{i6}\delta_{j3} - \delta_{i3}\delta_{j6},$$
$$2\tilde{P}_{ij}^1 = \delta_{i5}\delta_{j4} - \delta_{i4}\delta_{j5}, \quad 2\tilde{P}_{ij}^2 = \delta_{i6}\delta_{j4} - \delta_{i4}\delta_{j6} \quad (12)$$

for the Yukawa interactions $y_{U,D}$, and the vectorlike masses for the new hyperfermions via the matrix

$$M = \begin{pmatrix} m_1 i\sigma_2 & 0 & 0 \\ 0 & m_2 i\sigma_2 & 0 \\ 0 & 0 & -m_3 i\sigma_2 \end{pmatrix}. \quad (13)$$

Note that for $m_1 = m_2 = m_3$, the mass matrix is proportional to the EW-preserving vacuum in Eq. (6) with $\theta = \pi$ ($c_\theta = -1$): This is not by chance, as it is indeed the hyperfermion masses that determine the signs in the vacuum structure [42].

By demanding the left- and right-handed SM quarks and leptons transform as $q_{L,i} \equiv (u_{L,i}, d_{L,i})^T \to q_{L,i}$, $l_{L,i} \equiv (\nu_{L,i}, e_{L,i})^T \to l_{L,i}$, $u_{R,i} \to -u_{R,i}$, $d_{R,i} \to -d_{R,i}$, and $e_{R,i} \to -e_{R,i}$ under the $\mathbb{Z}_2$ symmetry, the elementary $\mathbb{Z}_2$-odd scalar doublet $H$ couples to the SM fermions like the Higgs doublet in the SM with the Yukawa interactions preserving the $\mathbb{Z}_2$ symmetry:

$$\mathcal{L}_Y = -y_t^{ij} \bar{q}_{L,i} H u_{R,j} - y_b^{ij} \bar{q}_{L,i} \tilde{H} d_{R,j}$$
$$- y_e^{ij} \bar{l}_{L,i} \tilde{H} e_{R,j} + \text{H.c.} \quad (14)$$

Below the condensation scale $\Lambda_{\text{HC}} \sim 4\pi f$, Eq. (11) yields the following leading-order effective potential:

$$V_{\text{eff}}^0 = m_H^2 H^\dagger H + \lambda_H (H^\dagger H)^2$$
$$- 4\pi f^3 Z \left(\frac{1}{2}\text{Tr}[M\Sigma] - y_U H_\alpha \text{Tr}[P^\alpha \Sigma]\right.$$
$$\left. - y_D \tilde{H}_\alpha \text{Tr}[\tilde{P}^\alpha \Sigma] + \text{H.c.}\right), \quad (15)$$

where $Z$ is a nonperturbative $\mathcal{O}(1)$ constant that can be suggested by lattice simulations [e.g., $Z \approx 1.5$ in Ref. [43] for the SU(2) gauge theory with two Dirac (four Weyl) hyperfermions]. To next-to-leading order, the EW gauge interactions contribute to the effective potential given by

$$V_{\text{gauge}}^{1-\text{loop}} = C_g f^4 \left(\sum_{i=1}^3 g_L^2 \text{Tr}[T_L^i \Sigma T_L^{iT} \Sigma^\dagger]\right.$$
$$\left. + g_Y^2 \text{Tr}[T_R^3 \Sigma T_R^{3T} \Sigma^\dagger]\right)$$
$$= -C_g f^4 \left(\frac{3g_L^2 + g_Y^2}{2} c_\theta^2 + \ldots\right), \quad (16)$$

where $C_g$ is an $\mathcal{O}(1)$ form factor that can be computed on the lattice, and the gauged generators embedded in the $SU(2)_L \otimes SU(2)_R$ subgroup of the global symmetry group $SU(6)$ are identified by the left and right generators

$$T_L^i = \begin{pmatrix} \sigma_i & 0 & 0 \\ 0 & 0 & 0 \\ 0 & 0 & \sigma_i \end{pmatrix}, \quad T_R^i = \begin{pmatrix} 0 & 0 & 0 \\ 0 & -\sigma_i^T & 0 \\ 0 & 0 & 0 \end{pmatrix} \quad (17)$$

with $i = 1, 2, 3$ and $\sigma_i$ as the Pauli matrices. This effective potential is at the one-loop level, and accordingly, the contribution is subleading comparable to the vectorlike mass terms. However, we will include this contribution in the following numerical calculations because these terms are essential for generating the small neutrino masses in this model, as discussed in Sec. VI. Thus, to the leading order, the effective potential of Eq. (15) as a function of the misalignment angle $\theta$ and the elementary field $\sigma_h$ in Eq. (10) reads

$$V_{\text{eff}} = \frac{1}{2} m_H^2 \sigma_h^2 + \frac{1}{4} \lambda_H \sigma_h^4$$
$$+ 8\pi f^3 Z \left(m_{23} c_\theta - \frac{y_{UD}}{\sqrt{2}} \sigma_h s_\theta\right), \quad (18)$$

where $m_{23} \equiv m_2 + m_3$ and $y_{UD} \equiv y_U + y_D$. By assuming $\langle \sigma_h \rangle = v$ in Eq. (10), the vacuum conditions then read

$$0 = \left.\frac{\partial V_{\text{eff}}}{\partial \sigma_h}\right|_{\sigma_h = v} = -4\sqrt{2}\pi Z y_{UD} f^3 s_\theta + m_\lambda^2 v,$$
$$0 = \left.\frac{\partial V_{\text{eff}}}{\partial \theta}\right|_{\sigma_h = v} = 8\pi Z f^3 \left(m_{23} s_\theta + \frac{y_{UD}}{\sqrt{2}} v c_\theta\right), \quad (19)$$

where $m_\lambda^2 \equiv m_H^2 + \lambda_H v^2$. The first condition visualizes that the VEV generated by the vacuum misalignment in the composite sector $f s_\theta$ can be transferred via the Yukawa interactions $y_{U,D}$ to the neutral CP-even component of $H$, leading to a VEV $v$ of $\sigma_h$. These vacuum conditions yield the parameter expressions

$$m_{23} = -\frac{c_\theta m_\lambda^2 t_\beta^2}{8\pi Z f}, \quad y_{UD} = \frac{t_\beta m_\lambda^2}{4\sqrt{2}\pi Z f^2} \quad (20)$$

with $t_\beta \equiv v/(f s_\theta)$. Unlike the Higgs mechanism in the SM, the squared mass parameter $m_H^2$ of the elementary Higgs doublet in Eq. (15) does not need to change the sign to trigger the EWSB.

According to the Higgs potential in Eq. (15), the neutral $CP$-even scalar mass matrix in the basis $(\sigma_h, h)$ can be written as

$$M^2_{\text{Higgs}} = m_\lambda^2 \begin{pmatrix} 1+\delta & -c_\theta t_\beta \\ -c_\theta t_\beta & t_\beta^2 \end{pmatrix}, \quad (21)$$

where $\delta \equiv 2\lambda_H v^2/m_\lambda^2$. The mass eigenstates are given in terms of the interaction eigenstates by

$$h_1 = c_\alpha \sigma_h - s_\alpha h, \qquad h_2 = s_\alpha \sigma_h + c_\alpha h \quad (22)$$

with

$$t_{2\alpha} = \frac{2t_\beta c_\theta}{1 - t_\beta^2 + \delta}. \quad (23)$$

The corresponding masses of these eigenstates are

$$m^2_{h_{1,2}} = \frac{m_\lambda^2}{2} [1 + t_\beta^2 + \delta \pm (2c_\theta t_\beta s_{2\alpha} + (1 - t_\beta^2 + \delta) c_{2\alpha})]. \quad (24)$$

The lightest of these eigenstates will be identified with the 125-GeV SM Higgs boson. For $1 + \delta < t_\beta^2$ and in the following, this is the $h_1$ state.

This model features seven parameters relevant to our study: $m_1$, $m_{23}$, $y_{UD}$, $m_H$, $\lambda_H$, $s_\theta$, $f$, and $t_\beta$, and four constraint equations including the EW scale in Eq. (3), the two vacuum conditions in Eq. (19), and the Higgs mass in Eq. (24). From now on, we set $m_1 = 0$ for simplicity. Thus, we can assume that $m_H$, $f$, and $t_\beta$ are free parameters. Furthermore, the SM naturalness problem is almost completely alleviated by fixing $\Lambda_{\text{HC}} \sim 4\pi f \equiv m_{\text{Planck}}$ and $m_H \equiv f \approx 10^{18}$ GeV, which requires a small vacuum misalignment of $s_\theta = 1 \times 10^{-16} \ldots 5 \times 10^{-18}$ for $t_\beta = 2 \ldots 50$. According to 't Hooft's naturalness principle [1], such a small misalignment is technically natural due to the restoration of the global $\mathbb{Z}_2$ symmetry when $s_\theta = 0$. From now on, the only free parameter is $t_\beta$.

## IV. RG ANALYSIS AND VACUUM STABILITY

As concluded in Ref. [16], PCH models may suffer from a low vacuum instability scale, which is the energy scale below the Planck scale where the quartic coupling, e.g., $\lambda_H$ in Eq. (11), runs to negative values. Such an instability originates from the fact that the top Yukawa coupling of the elementary interaction eigenstate $\sigma_h$ is enhanced compared to the SM Higgs boson [44] by

$$y_t = y_t^{\text{SM}}/s_\beta, \quad (25)$$

where $y_t$ and $y_t^{\text{SM}}$ are the enhanced and SM top Yukawa couplings, respectively.

In the following, we numerically calculate the RG running of the quartic coupling $\lambda_H$ in Eq. (11). Moreover, in this RG analysis, we assume that the top Yukawa and the EW gauge couplings run as in the SM below the compositeness scale, since all values of the couplings are SM-like after the condensation, and we neglect the effects on the quartic coupling running during the confinement. We will, therefore, use the SM RG equations up to three loops in Ref. [45]. We fix the initial values of the SM gauge couplings at $\mu_Z = m_Z$, and the value of the SM top Yukawa from $y_t^{\text{SM}}(m_t)$ as in Ref. [46]. We assume further that no composite states other than the Higgs state affect the running of $\lambda_H$ due to the fact that all of these states are integrated out below $f \approx 1 \times 10^{18}$ GeV, because their masses are between $f$ and $\sim ft_\beta$ for small $t_\beta^{-2}$. For example, expanding in $s_\theta^2$ and $t_\beta^{-2}$, the mass eigenstate $h_2$ in Eq. (22) has the mass $m_{h_2} \simeq f[t_\beta + 1/(2t_\beta)]$.

By requiring stability of the vacuum up to the compositeness scale, an upper bound on the mass parameter is found in Ref. [16] to be roughly $m_H^2 \lesssim (100 s_\theta^{-1} \text{ GeV})^2$. In our scenario, where the mass parameter is close to the Planck scale, it is impossible to avoid a small vacuum instability scale without decreasing $s_\theta$ to a small value. Fortunately, a small value of $s_\theta$ is technically natural in this model. Therefore, we can avoid a critical small instability scale, even if $m_H = f \approx 1 \times 10^{18}$ GeV, requiring that $s_\theta \lesssim 1 \times 10^{-16}$ ($t_\beta \gtrsim 2$). This is illustrated in Fig. 1, where different RG evolutions of the quartic coupling $\lambda_H$ for various $t_\beta$ are plotted by varying the RG scale, $\mu$. The black dashed curve represents the RG evolution of the SM quartic coupling, $\lambda_{\text{SM}}$.

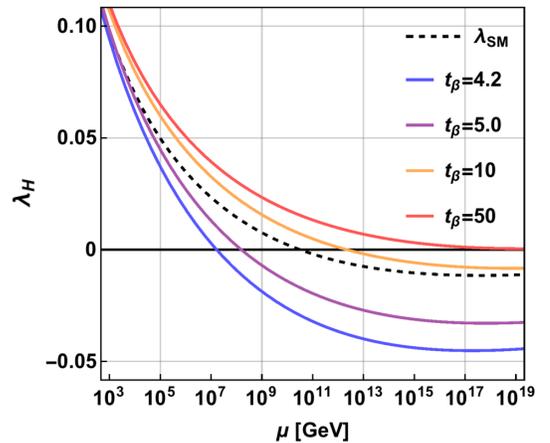

FIG. 1. RG evolution of the quartic coupling $\lambda_H$ up to three loops as a function of the RG scale $\mu$ for various $t_\beta = 4.2, 5.0, 10, 50$, where $C_g = 1.0$ and $m_H = f \approx 1 \times 10^{18}$ GeV. The dashed black line represents the RG evolution of the SM quartic coupling $\lambda_{\text{SM}}$. In all of these RG calculations, we have used the best measured values of the different masses and couplings. For example, $m_h = 125.25 \pm 0.17$ GeV and $m_t = 172.76 \pm 0.30$ GeV [47].

The figure shows that the vacuum instability scale increases for larger $t_\beta$. According to the figure, the estimate of the upper bound of $m_H$ in Ref. [16] is slightly inaccurate for large compositeness scales in this model; for example, we need that $t_\beta \gtrsim 50$ ($s_\theta \lesssim 5 \times 10^{-18}$) to obtain a stable vacuum up to the Planck scale when $m_H = f \approx 1 \times 10^{18}$ GeV.

However, an intriguing outcome of the Higgs discovery has been the finding that the certain Higgs mass leads to a vacuum lying very close to the boundary between the stability and metastability regions [45]. Although the SM quartic coupling runs to negative values at $\sim 10^{10}$ GeV as shown by the dashed curve in Fig. 1, the SM vacuum may be metastable and thus not unstable. The same is true of the vacuum in this model. In Fig. 4 in Ref. [45], this near criticality of the Higgs boson is investigated and depicted in a phase diagram in terms of the SM quartic Higgs coupling $\lambda_{SM}$ and top Yukawa coupling $y_t^{SM}$ normalized at the Planck scale. A similar stability phase diagram for this model is shown here in Fig. 2 in terms of $t_\beta$ and $\lambda_H$ renormalized at the Planck scale, where the green, yellow, and red regions represent, respectively, a stable, metastable, and unstable vacuum. For the entire phase diagram, the top Yukawa coupling at the Planck scale is in the range $y_t(m_{\text{Planck}}) = 0.40...0.53$, giving rise to a metastability constraint of the quartic coupling of $\lambda_H(m_{\text{Planck}}) \gtrsim -0.046$ [45] and stability constraint $\lambda_H(m_{\text{Planck}}) \gtrsim 0$.

In Fig. 2, the red (blue) curves show the stability state of the vacuum for $C_g = 1.0$ ($C_g = 5.0$) with $3\sigma$ band of the top mass $m_t = 172.76 \pm 0.30$ GeV, where $C_g$ is the $\mathcal{O}(1)$ form factor of the gauge loop potential contributions in Eq. (16). For $C_g = 1.0$, the vacuum may be minimum metastable down to $t_\beta \approx 4.1$ ($t_\beta \approx 3.9$ in the $3\sigma$ band of the top mass), while there may be no instability constraint of $t_\beta$ for $C_g = 5.0$. However, there are no solutions for the Higgs mass condition for $t_\beta < 2.1$. Thus, the vacuum with the values $t_\beta = 4.2$ and $C_g = 1.0$ lies close to the boundary between metastability and instability, even though the quartic coupling runs to negative values already at $\sim 10^7$ GeV, as shown by the blue curve in Fig. 1. For large $t_\beta$, the subleading contributions from the EW gauge interactions in Eq. (16) are neglectable in the effective potential shown by the fact that the red ($C_g = 1.0$) and blue ($C_g = 5.0$) curves merge in Fig. 2 for $t_\beta \gtrsim 10$. Finally, for $t_\beta \gtrsim 13.5$, the vacuum may be stable for the lower part of the $3\sigma$ band of the top mass.

In conclusion, if we permit the vacuum to be metastable like in the SM, it allows lower values of $t_\beta$. As we will see in Sec. VI, smaller values of $t_\beta$ are necessary for this model to generate viable neutrino masses and mixing. Before considering the neutrino physics of this model, we will consider how the parameter space can be searched by gravitational waves from a confinement-induced phase transition.

## V. CONSTRAINTS FROM GRAVITATIONAL WAVES

This theory undergoes a phase transition, where the new strong dynamics confines at about the temperature $T_* \approx f$. This phase transition may generate gravitational waves (GWs) [48]. During a first-order phase transition, GWs can be generated by bubble collisions [49,50], stirred acoustic waves [51,52], and magnetohydrodynamic turbulence in the supercooling plasma [53,54]. The total power spectrum of the GWs consists of these three components, which can be written as

$$h^2\Omega(\nu) = h^2\Omega_{\text{env}}(\nu) + h^2\Omega_{\text{sw}}(\nu) + h^2\Omega_{\text{turb}}(\nu), \quad (26)$$

where $\nu$ denotes the frequency of the GWs, while $h^2\Omega_{\text{env}}(\nu)$, $h^2\Omega_{\text{sw}}(\nu)$, and $h^2\Omega_{\text{turb}}(\nu)$ are the power spectra of the GWs, respectively, from bubble collisions in the envelope approximation (found in Ref. [49]), acoustic waves (in Ref. [51]), and Kolmogorov-type turbulence (in Refs. [53,54]). The full expressions of these spectra are given in Ref. [55]. The GW spectrum from first-order phase transitions is generally characterized by two essential parameters $\alpha(T_*)$ and $\beta(T_*)/H_*$ evaluated at the temperature during the phase transition, $T_* \approx f$. The parameter $\alpha$ is the ratio of the vacuum energy density and radiation energy density, while $\beta/H_*$ is the nucleation rate divided with the Hubble rate during the phase transition, which measures the time duration of the phase transition.

In Fig. 3, the present (upper) and future (lower panel) exclusions in the parameter space of $\alpha$ and $\beta/H_*$ are depicted for various compositeness scales $\Lambda_{\text{HC}} = m_{\text{Planck}}$ (blue), $10^{17}$ GeV (red), and $10^{15}$ GeV (green regions) set by the constraints from Refs. [56,57] on $h^2\Omega$ in Eq. (26) (the full expression of $h^2\Omega$ is given in Ref. [55]). In Ref. [56], the first experimental upper limits on the presence of stochastic GWs are in a frequency band above 1 THz set by an experimental setup of graviton-to-photon conversion in a constant magnetic field. Those constraints exclude GWs in the frequency bands from $(2.7 - 14) \times 10^{14}$ Hz and $(5 - 12) \times 10^{18}$ Hz down to amplitudes $h^2\Omega \approx 6 \times 10^{-26}$ and $h^2\Omega \approx 5 \times 10^{-28}$ at 95% confidence level, respectively. Furthermore, according to Ref. [57], GHz GWs can be probed by a graviton-magnon detector which measures the resonance fluorescence of magnons. The sensitivity of this detector reaches amplitudes at $h^2\Omega \sim 10^{-19}$ and $h^2\Omega \sim 10^{-21}$ with the frequencies 14 and 8.2 GHz, respectively. However, that detector sets typically weaker constraints of this model than the THz detectors in Ref. [56] due to their weaker sensitivities and the fact that the peak frequencies of the GWs produced by this model are larger relative to the GHz ballpark. The peak frequency of the GWs for $\Lambda_{\rm HC} = m_{\rm Planck}$ is approximately given by $\nu_{\rm peak} \approx 3 \times 10^{11}(\beta/H_*)$ Hz and is, therefore, typically larger relative to the GHz range.

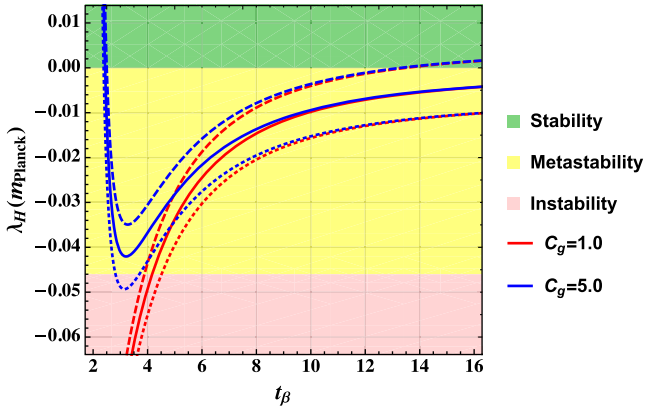

FIG. 2. The phase diagram of the vacuum stability in terms of $t_\beta$ and the quartic coupling renormalized at the Planck scale $\lambda_H(m_{\text{Planck}})$. Depending on these parameters, the vacuum is either stable, metastable, or unstable. The red (blue) lines show the stability state of the vacuum for $C_g = 1.0$ ($C_g = 5.0$) in the $3\sigma$ band of the top mass $m_t = 171.86...173.66$ GeV [47] from the dashed to dotted lines, while for the solid lines, the top mass is at its expectation value of $m_t = 172.76$ GeV [47].

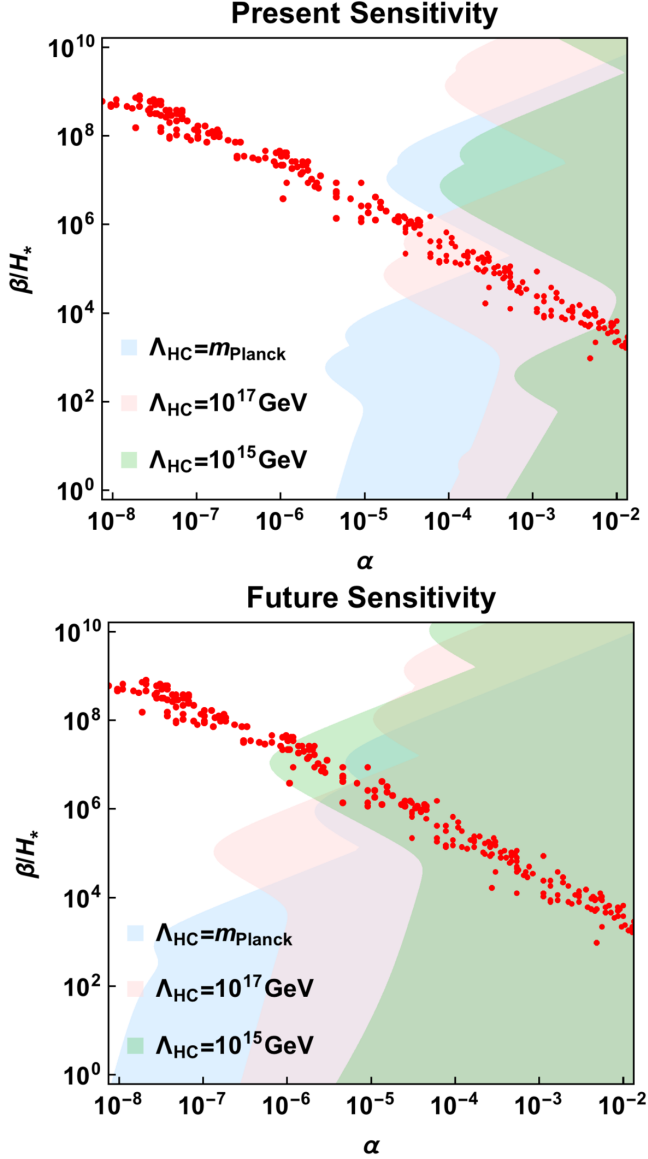

FIG. 3. The present (upper panel) and future (lower panel) exclusions are depicted in the parameter space of $\alpha$ and $\beta/H_*$ for various compositeness scales $\Lambda_{\rm HC} = m_{\rm Planck}$ (blue), $10^{17}$ GeV (red), and $10^{15}$ GeV (green shaded region) set by the GW experiments in Refs. [56,57]. The red points give successful SFOEWPTs.

In the following, we will make an analysis of the parameter space of $\alpha$ and $\beta/H_*$ in this model leading to successful strong first-order EW phase transitions (SFOEWPTs). A similar analysis is made in Ref. [58], where the CH coset $SO(6)/SO(5) \sim SU(4)/Sp(4)$ is considered. At finite temperature, the effective Lagrangian of the composite PNGBs $h$ and $\eta$ in Sec. III B receives thermal corrections and vacuum structure changes. To the leading $T^2$ terms, the finite temperature effective potential in terms of $h$ and $\eta$ is given by [59]

$$V_T(h,\eta) = \frac{\mu_h^2 + c_h T^2}{2} h^2 + \frac{\lambda_h}{4} h^4 + \frac{\mu_\eta^2 + c_\eta T^2}{2} \eta^2 + \frac{\lambda_\eta}{4} \eta^4 + \frac{\lambda_{h\eta}}{2} h^2\eta^2, \quad (27)$$

where

$$c_h = \frac{3g_L^2 + g_Y^2}{16} + \frac{\lambda_h}{2} + \frac{\lambda_{h\eta}}{12}, \qquad c_\eta = \frac{\lambda_\eta}{4} + \frac{\lambda_{h\eta}}{3}.$$

The coefficients $\mu_{h,\eta}^2$ and $\lambda_{h,\eta,h\eta}$ are determined by the vectorlike masses of the hyperfermions in Eq. (13), the gauge interactions in Eq. (16), and the mass mixing between $h$ and $\sigma_h$ in Eq. (21).

The necessary condition for SFOEWPT is the existence of two degenerate vacua at some critical temperature $T_c$. Here, the red points in Fig. 3 fulfill the conditions for a "two-step" phase transition[2] in which the VEVs of the composite PNGBs $h$ and $\eta$ ($\langle h \rangle, \langle \eta \rangle$) changed as $(0,0) \to (0, v_\eta) \to (v_h, 0)$ with $v_h = f s_\theta$ when the Universe was cooled down from the temperature $T \gg m_h$ to $T = 0$. Therefore, there should exist two degenerate vacua at some critical temperature [58]

---

[2]There exists another possible SFOEWPT mechanism called the "one-step" SFOEWPT, in which a potential barrier is induced only along the $h$ direction, and the $\eta$ never achieves a VEV [59]. However, in this case, we need to include the thermal corrections depending linearly on T which leads to gauge-dependent critical temperature and VEV $T_c$ and $v_c$ [60]. Thus, we will not consider this scenario here.

$$T_c^2 = \frac{\mu_h^2\sqrt{\lambda_\eta} - \mu_\eta^2\sqrt{\lambda_h}}{c_\eta\sqrt{\lambda_h} - c_h\sqrt{\lambda_\eta}}. \quad (28)$$

The condition of two degenerate vacua for $V_T(h,\eta)$ in Eq. (27) is [58]

$$\frac{c_\eta}{c_h} < \frac{\mu_\eta^2}{\mu_h^2} < \frac{\sqrt{\lambda_\eta}}{\sqrt{\lambda_h}} < \frac{\lambda_{h\eta}}{\lambda_h}. \quad (29)$$

Note that degenerate vacua are necessary but not adequate for a SFOEWPT. To achieve a SFOEWPT, the critical condition

$$\frac{S_3(T_n)}{T_n} \sim 4\ln\left(\frac{\xi m_{\text{Planck}}}{T_n}\right) \sim 140 \quad (30)$$

should be satisfied at some nucleation temperature $T_n$, which can be checked by calculating the bubble nucleation rate per volume in the early Universe

$$\frac{\Gamma}{V} \approx T^4 \left(\frac{S_3}{2\pi T}\right)^{3/2} e^{-S_3(T)/T}. \quad (31)$$

Here, $S_3$ is the classical action of the $O(3)$ symmetric bounce solution [61] and $\xi \approx 0.03$.

Numerically, we use the CosmoTransitions package [62] to estimate the parameters $\alpha$ and $\beta/H_*$ that provide degenerate vacua and fulfill the condition in Eq. (30), leading to SFOEWPTs. These parameter sets are represented by the red points in Fig. 3. Thus, regarding the figure for $\Lambda_{\text{HC}} = m_{\text{Planck}}$, there is an upper limit of the parameter $\alpha \lesssim 3 \times 10^{-4}$ and lower limit of $\beta/H_* \gtrsim 1 \times 10^5$, which may be improved to $\alpha \lesssim 8 \times 10^{-6}$ and $\beta/H_* \gtrsim 4 \times 10^6$ by the proposed future experiments in Ref. [56]. Overall, it opens up the possibility of probing this model with GWs by improving the sensitivities in the GW experiments in the frequency band above 1 THz.

## VI. LOOP-INDUCED NEUTRINO MASSES

Another mysterious missing piece in our understanding of the Universe is that the neutrinos have masses that are several orders of magnitude smaller than those of the charged fermions. The simplest solution to this puzzle is the seesaw mechanism [63–65], which requires a typical new scale $\Lambda_{\text{seesaw}} \approx 10^{12}$ GeV. In this section, we explore the possibility of realizing a one-loop radiative seesaw mechanism in this partially composite multi-Higgs scheme by considering the studies in Ref. [41]. However, the neutral component of a $\mathbb{Z}_2$-even doublet, not of a $\mathbb{Z}_2$-odd as in Ref. [66], runs in the loop as in Fig. 4.

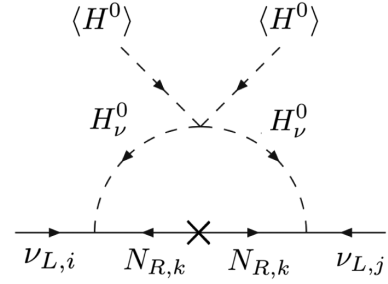

FIG. 4. One-loop radiative Majorana neutrino masses in this model similar to the scotogenic model proposed in Ref. [66].

### A. The Lagrangian and scalar potential terms in the neutrino sector

To incorporate this mechanism, we need to add a new $\mathbb{Z}_2$-even elementary weak doublet with hypercharge $+1/2$,

$$H_\nu = \frac{1}{\sqrt{2}}\begin{pmatrix} \sigma_R - i\sigma_I \\ -(\pi_\nu^2 + i\pi_\nu^1) \end{pmatrix}, \quad (32)$$

and three right-handed neutrino SM singlets $N_{R,i}$ with $i = 1, 2, 3$ transforming even under the global $\mathbb{Z}_2$ symmetry.

Thus, new fundamental Yukawa couplings with the neutrino fields can be written as

$$\mathcal{L}_Y \supset -h^{ij}\bar{l}_{L,i}H_\nu N_{R,j} + \text{H.c.} \quad (33)$$

Furthermore, the new elementary $\mathbb{Z}_2$-even doublet $H_\nu$ is allowed to be coupled to the composite Higgs sector by adding new fundamental Yukawa couplings between $H_\nu$ and the $\mathbb{Z}_2$-even hyperfermions to Eq. (11):

$$\begin{aligned}\mathcal{L}_{\text{PCH}} \supset &-y_1 H_{\nu,\alpha}(\Psi^T P_\nu^\alpha \Psi) \\ &- y_2 \tilde{H}_{\nu,\alpha}(\Psi^T \tilde{P}_\nu^\alpha \Psi) + \text{H.c.},\end{aligned} \quad (34)$$

where we introduce the spurions

$$\begin{aligned}2P_{\nu,ij}^1 &= \delta_{i1}\delta_{j3} - \delta_{i3}\delta_{j1}, & 2P_{\nu,ij}^2 &= \delta_{i2}\delta_{j3} - \delta_{i3}\delta_{j2}, \\ 2\tilde{P}_{\nu,ij}^1 &= \delta_{i1}\delta_{j4} - \delta_{i4}\delta_{j1}, & 2\tilde{P}_{\nu,ij}^2 &= \delta_{i2}\delta_{j4} - \delta_{i4}\delta_{j2}.\end{aligned} \quad (35)$$

In the following, we assume $y_1 = y_2$ for simplicity. Below the condensation scale $\Lambda_{\text{HC}} \sim 4\pi f$, Eq. (34) yields the following contributions to the effective potential in Eq. (15):

$$\begin{aligned}V_{\text{eff}}^0 \supset 4\pi f^3 Z_\nu (&y_1 H_{\nu,\alpha}\text{Tr}[P_\nu^\alpha \Sigma] \\ &+ y_2 \tilde{H}_{\nu,\alpha}\text{Tr}[\tilde{P}_\nu^\alpha \Sigma] + \text{H.c.}),\end{aligned} \quad (36)$$

where $Z_\nu$ is a nonperturbative $\mathcal{O}(1)$ constant. From now on, we assume that $Z_\nu \equiv Z \approx 1.5$ for simplicity.

Finally, we add all the allowed terms including the new doublet to the underlying Lagrangian in Eq. (11) that conserves all the symmetries, which are given by

$$\mathcal{L}_{\text{PCH}} \supset D_\mu H_\nu^\dagger D^\mu H_\nu - m_{H_\nu}^2 H_\nu^\dagger H_\nu \\ - \lambda_{H_\nu}(H_\nu^\dagger H_\nu)^2 - \lambda_1 H^\dagger H H_\nu^\dagger H_\nu \\ - \lambda_2 H^\dagger H_\nu H_\nu^\dagger H - (\lambda_3 (H^\dagger H_\nu)^2 + \text{H.c.}). \quad (37)$$

In the following, we assume $m_{H_\nu} \equiv m_H = f$ to avoid naturalness problems and $\lambda_{1,2,3} = 1$ for simplicity, which has an insignificant influence on the RG running of $\lambda_H$ since the components of $H_\nu$ have large $\mathcal{O}(f)$ masses and are "integrated out." Note that the above potential contributions from the neutrino sector do not change the masses of the mass eigenstates $h_{1,2}$ in Eq. (24), where $h_1$ is still identified with the 125-GeV SM Higgs boson. Furthermore, the new elementary doublet has no influence on the vacuum structure discussed in Sec. III.

### B. The masses and mixing of the neutrinos

The mixing mass matrices $M_R^2$ and $M_I^2$ in the bases $(\sigma_R, \phi_R)$ and $(\sigma_I, \phi_I)$, respectively, are generated by the potential of the neutral components of the elementary and composite $\mathbb{Z}_2$-even Higgs doublets in Eqs. (9) and (32), which are given by

$$M_R^2 = \tilde{M}^2 + \begin{pmatrix} 2\lambda_3 v^2 & 0 \\ 0 & \frac{1}{2}C_g(3g_L^2 + g_Y^2)f^2c_\theta \end{pmatrix},$$

$$M_I^2 = \tilde{M}^2 + \begin{pmatrix} -2\lambda_3 v^2 & 0 \\ 0 & \frac{1}{2}C_g(8g_L^2 + (g_Y^2 - 5g_L^2)c_\theta)f^2 \end{pmatrix}$$
(38)

with

$$\tilde{M}^2 \equiv \begin{pmatrix} m_{H_\nu}^2 + \frac{1}{2}(\lambda_1 + \lambda_2)v^2 & -4\sqrt{2}\pi Z y_{12} f^2 c_{\theta/2} \\ -4\sqrt{2}\pi Z y_{12} f^2 c_{\theta/2} & 8\pi Z f m_1 + \frac{1}{2}m_\lambda^2 t_\beta^2 \end{pmatrix},$$

where $y_{12} \equiv y_1 + y_2$. Therefore, a mass splitting is generated between the masses $m_{R,I}$ of the mass eigenstates $\tilde{\sigma}_{R,I}$ consisting mostly of the neutral components $\sigma_{R,I}$ in Eq. (32), respectively.

Assuming that the right-handed neutrinos $N_{R,i}$ are not much heavier than the neutral components of the new $\mathbb{Z}_2$-even doublet, small Majorana masses of the left-handed neutrinos are generated by the loop diagram shown in Fig. 4, analogous to the one in the traditional scotogenic model [66]. The loop diagram results in the mass expression [66]

$$m_\nu^{ij} = \sum_{k=1}^{3} \frac{h^{ik}h^{jk}}{(4\pi)^2} M_k \left[ \frac{m_R^2}{m_R^2 - M_k^2} \ln\left(\frac{m_R^2}{M_k^2}\right) \\ - \frac{m_I^2}{m_I^2 - M_k^2} \ln\left(\frac{m_I^2}{M_k^2}\right) \right] \equiv \sum_{k=1}^{3} h^{ik}h^{jk} \Xi_{\nu,k}, \quad (39)$$

where $M_i$ denotes the masses of the right-handed neutrinos, $N_{R,i}$. To obtain nonzero neutrino masses, we need a mass splitting between the masses $m_{R,I}$ of the mass eigenstates $\tilde{\sigma}_{R,I}$, respectively, given by Eq. (38).

By ignoring the gauge loop potential contributions in Eq. (38), i.e., $g_{L,Y} = 0$, the mass splitting between the mass eigenstates $\tilde{\sigma}_{R,I}$ will only depend on the $\lambda_3 v^2$ term, which is negligible relative to their masses since $m_{R,I} \sim f \gg v$. In this scenario, either the neutrino Yukawa couplings $h^{ij}$ in Eq. (33) or the quartic coupling $\lambda_3$ in Eq. (37) needs to be nonperturbatively large ($h^{ij}, \lambda_3 > 4\pi$) to achieve large enough neutrino masses. When the gauge interactions are turned on, a mass splitting will be generated in the order of $f$ between the composite states $\phi_{R,I}$, resulting in a more significant mass splitting of $\tilde{\sigma}_{R,I}$ via the Yukawa couplings $y_{1,2}$ in Eq. (34). As shown in the following, this mass splitting may be enough to generate large enough neutrino masses with perturbative couplings. In this model, the neutrino masses are thus dynamically loop induced by the composite dynamics via the EW gauge and Yukawa interactions.

Before we present the numerical calculations, we need to define the neutrino mass matrix in Eq. (39), which can be diagonalized as

$$m_\nu^{\text{Diag}} = U_{\text{PMNS}}^T m_\nu U_{\text{PMNS}} = \text{Diag}(m_{\nu_1}, m_{\nu_2}, m_{\nu_3}), \quad (40)$$

where $m_{\nu_i}$ with $i = 1, 2, 3$ are the left-handed neutrino masses. The matrix $U_{\text{PMNS}} = UU_m$ is the Pontecorvo-Maki-Nakagawa-Sakata (PMNS) matrix, where $U_m = \text{Diag}(1, e^{i\phi_1/2}, e^{i\phi_2/2})$ encoding the Majorana phases and the matrix $U$ is parametrized as

$$\begin{pmatrix} c_{12}c_{13} & s_{12}c_{13} & s_{13}e^{-i\delta} \\ -s_{12}c_{23} - c_{12}s_{23}s_{13}e^{i\delta} & c_{12}c_{23} - s_{12}s_{23}s_{13}e^{i\delta} & s_{23}c_{13} \\ s_{12}s_{23} - c_{12}c_{23}s_{13}e^{i\delta} & -c_{12}s_{23} - s_{12}c_{23}s_{13}e^{i\delta} & c_{23}c_{13} \end{pmatrix}$$

with the mixing angles $s_{ij} \equiv \sin\theta_{ij}, c_{ij} \equiv \cos\theta_{ij}$, and the Dirac phase $\delta$. In this paper, we assume that the Majorana and Dirac phases are vanishing ($\phi_{1,2} = 0$ and $\delta = 0$), but it is possible to add them without significant changes of our conclusions.

### C. Numerical results

In the following, we fit to the best-fit experimental values for the mass-squared differences ($\Delta m_{ij} \equiv m_{\nu_i}^2 - m_{\nu_j}^2$) and mixing angles ($s_{ij} \equiv \sin\theta_{ij}$), which are given in Ref. [67]

for both normal hierarchy (NH) of the neutrinos, i.e., $m_{\nu_1} < m_{\nu_2} < m_{\nu_3}$, and inverted hierarchy (IH), i.e., $m_{\nu_3} < m_{\nu_1} < m_{\nu_2}$. Finally, we also include the upper bound on the sum of the neutrino masses coming from cosmology. The most reliable bound is from the Planck Collaboration [68],

$$\sum_i m_{\nu_i} \lesssim 0.23 \text{ eV}. \quad (41)$$

In the following calculations, we have chosen the NH of the neutrinos, where $m_{\nu_1} = 0.0010$ eV, leading to $m_{\nu_2} = 0.0087$ eV, $m_{\nu_3} = 0.050$ eV, and therefore, the bound in Eq. (41) is fulfilled. If we choose IH or another value of $m_{\nu_1}$, the following conclusions will not change significantly. Furthermore, we assume that the right-handed neutrino masses are $M_i/2 = m_H = m_{H_\nu} = f \approx 1 \times 10^{18}$ GeV and the vectorlike mass of the hyperfermion doublet $\Psi_1$ is $m_1 = m_{2,3}/2$, because those values give rise to the most favorable results.

In Fig. 5, for $\Lambda_{HC} \sim 4\pi f = m_{Planck}$, we have plotted $\Xi_\nu \equiv \Sigma_{i=1}^3 \Xi_{\nu,i}$ [defined in Eq. (39)] as function of $t_\beta$ for various Yukawa couplings $y_{1,2}$ and for $C_g = 1.0, 2.5, 4.0, 5.0$. The black solid line represents the case with neutrino Yukawa couplings $h^{ij}$ of order unity resulting in viable neutrino masses and mixing in Eq. (40), while the black dashed line represents the case with the largest perturbative value of $h^{ij}$, i.e., below this line $\max(|h^{ij}|) > 4\pi$. Note that the red and green lines stop below $t_\beta \approx 3.9$ and $t_\beta \approx 3.7$, respectively, because the vacuum is unstable below these values with the top mass in its $3\sigma$ band, for example, shown for $C_g = 1.0$ (the red lines) in Fig. 2.

Assuming $C_g = 1.0$, $y_{1,2} = 10$, and $t_\beta = 3.9$, there exists one positive, real solution of the neutrino Yukawa coupling constants with maximal number of zeroes by using Eq. (39):

$$h^{ij} = \begin{pmatrix} 2.53 & 3.69 & 0.62 \\ 0 & 7.43 & 9.17 \\ 0 & 0 & 9.99 \end{pmatrix}. \quad (42)$$

Considering the upper limit of the $3\sigma$ band of the top mass for $C_g = 1.0$ (the dotted red line) in Fig. 2, it is required that $t_\beta \gtrsim 4.6$ to obtain a metastable vacuum. These values of $t_\beta$ may be too large to provide perturbative Yukawa couplings, $h^{ij}, y_{1,2} < 4\pi$, since the dashed red line (for $C_g = 1.0$ and $y_{1,2} = 4\pi$) in Fig. 5 is below the dashed black curve (i.e., $\max(|h^{ij}|) > 4\pi$) for $t_\beta \gtrsim 4.7$. Therefore, if $C_g$ is calculated by lattice simulations to be below 1.0, a more accurate measurement of the top mass may probe this theory.

The Yukawa couplings $h^{ij}, y_{1,2} \sim \mathcal{O}(1)$ by assuming that $C_g = 4.0$, $y_{1,2} = 5.0$ and $t_\beta = 3.2$. Thus, there exists one positive, real solution of the neutrino Yukawa coupling constants with maximal number of zeroes:

$$h^{ij} = \begin{pmatrix} 1.19 & 1.71 & 0.32 \\ 0 & 3.65 & 4.36 \\ 0 & 0 & 4.82 \end{pmatrix}. \quad (43)$$

Therefore, if the Yukawa couplings $h^{ij}, y_{1,2}$ are of order unity, we need that $C_g \gtrsim 4$ to obtain heavy enough neutrinos, where smaller Yukawa couplings require larger $C_g$. This theory can thus be probed by calculating $C_g$ by lattice simulations. In addition, there is a spectrum of solutions for $1.0 < C_g < 4.0$, where $\mathcal{O}(1) < h^{ij}, y_{1,2} < 10$.

There is also the possibility that the compositeness scale is smaller than the Planck scale, for example, $\Lambda_{HC} \sim 4\pi f = 10^{17}$ GeV with $M_i/2 = m_H = m_{H_\nu} = f \approx 8 \times 10^{15}$ GeV. Assuming that $C_g = 1.0$, $y_{1,2} = 2.5$, and $t_\beta = 4.2$, there exists one positive, real solution of the neutrino Yukawa coupling constants with maximal number of zeroes:

$$h^{ij} = \begin{pmatrix} 1.15 & 1.66 & 0.31 \\ 0 & 3.54 & 4.23 \\ 0 & 0 & 4.68 \end{pmatrix}. \quad (44)$$

With this compositeness scale, many solutions exist like this, where all the Yukawa couplings $h^{ij}, y_{1,2} \sim \mathcal{O}(1)$. However, we suppose in these calculations that the vacuum is metastable by assuming that the running of $\lambda_H$ above $\Lambda_{HC}$ does not result in an unstable vacuum. We have relegated the detailed study of the stability for the scenarios with $\Lambda_{HC} < m_{Planck}$ to future work.

Note for $\Lambda_{HC} \sim m_{Planck}$, all the dimensionful fundamental parameters of this theory are approximate of the order of $f \approx 1 \times 10^{18}$ GeV (close to the reduced Planck scale,

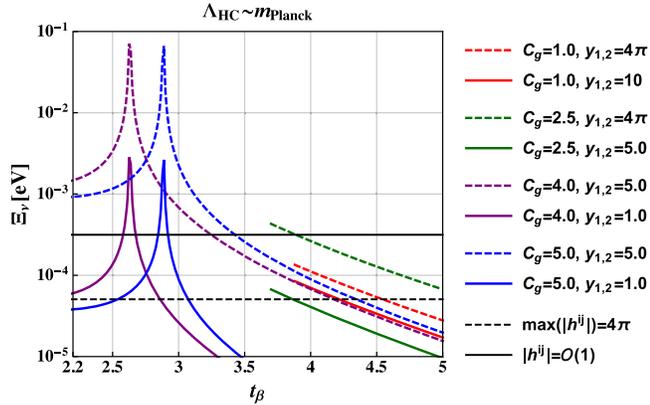

FIG. 5. $\Xi_\nu \equiv \Sigma_{i=1}^3 \Xi_{\nu,i}$ defined in Eq. (39) for varying $t_\beta$ for various Yukawa couplings $y_{1,2}$ and for $C_g = 1.0, 2.5, 4.0, 5.0$, where $\Lambda_{HC} \sim 4\pi f = m_{Planck}$. The black solid (or dashed) line represents the case with neutrino Yukawa couplings $h^{ij}$ of order unity [or with the largest perturbative value of $h^{ij}$, i.e., $\max(|h^{ij}|) = 4\pi$], resulting in viable neutrino masses and mixing in Eq. (40).

$\bar{m}_{\text{Planck}} \equiv m_{\text{Planck}}/\sqrt{8\pi} \approx 2.5f$): $m_H = m_{H_\nu} = M_i/2 = f$ and $6m_1 = 3m_{2,3} \approx f$. Furthermore, the reason for the choice of loop-induced neutrino masses instead of masses generated by a standard seesaw mechanism is the fact that the partially composite nature of the scalars appearing in the loop plays a crucial role in predicting a near-degenerate spectrum, which yields a large suppression of the loop-induced neutrino masses. Because of this large suppression, the right-handed neutrino masses up to the order of $f \approx 1 \times 10^{18}$ GeV can induce large enough neutrino masses. At the same time, a composite realization of the standard seesaw mechanism only requires right-handed neutrino masses, which are $M_i \lesssim \mathcal{O}(10^{15})$ GeV [69,70]. Therefore, the one-loop-induced neutrino masses allow right-handed neutrino masses of the order of $f \approx 1 \times 10^{18}$ GeV like the other dimensionful fundamental parameters of this theory.

## VII. FUTURE WORK

A future perspective of this model may be to extend it to dynamically generate the explicit masses of the fermions, like in the PCH model proposed in Ref. [15]. In such a model, the elementary scalar doublets $H$ and $H_\nu$ are extended to a complete two-index antisymmetric SU(6) representation, allowing for the Yukawa interactions of the elementary scalars and the hyperfermions to remain SU(6) symmetric. In such a model, the vectorlike masses $m_{1,2,3}$ and the right-handed neutrino masses $M_i$ may be generated via dynamically induced VEVs in the order of $f$ of the EW-singlet components of the new elementary scalar multiplet.

Finally, in this scenario with a compositeness scale close to the Planck scale, an interesting question arises about the global symmetries that determine the novel mechanism in this model. There are general expectations that all global symmetries are explicitly broken by gravitational effects [71], supported by theoretical calculations indicating explicit breaking of global symmetries by sources such as black holes [72] or wormholes [73]. However, there is so far no clear understanding of what the actual sources, mechanisms, or magnitude of the explicit breaking of global symmetries by gravity might be. For example, in many theories which are asymptotically safe, there are some indications that global symmetries might be preserved [74–79]. Furthermore, if there are violations of the global symmetries in this model their magnitude may be constrained by the cosmic birefringence measurements studied in Ref. [80]. The studies of such violations in this model are left for future work.

## VIII. CONCLUSIONS

This paper has presented a novel mechanism that may assist in UV completing (partially) composite Higgs models. With this mechanism, the compositeness scale of these models can be pushed up to the Planck scale in a technically natural manner, resulting in a complete remedy of the SM naturalness problem. For concreteness, we have demonstrated it in a minimal partially composite two-Higgs scheme with the composite coset SU(6)/Sp(6) featuring a $\mathbb{Z}_2$-odd and -even partially composite Higgs doublet as PNGBs. In this concrete model example, the $\mathbb{Z}_2$-odd scalar doublet triggers both the EWSB and the mass generation of the charged SM fermions based on vacuum misalignment with a small angle ($\sin\theta \lesssim 10^{-16}$), leading to softly breaking the $\mathbb{Z}_2$ symmetry of the composite dynamics. According to the 't Hooft naturalness principle [1], this limit ($\sin\theta \to 0$) is technically natural as it reveals the restoration of the global $\mathbb{Z}_2$ symmetry. Furthermore, a natural near degeneracy of the neutral components of the $\mathbb{Z}_2$-even scalar doublet features small loop-induced neutrino masses, where the composite dynamics generates this nondegeneracy via the EW gauge and Yukawa interactions.

A vacuum stability analysis of this concrete partially composite Higgs model results in a lower bound on the parameter $t_\beta$ that provides either a metastable or stable vacuum. This lower bound depends on the nonperturbative coefficient of the EW gauge loop potential contributions $C_g$ and the top mass. For example, when $C_g = 1.0$, the vacuum may be minimum metastable down to $t_\beta \approx 3.9$ in the $3\sigma$ band of the top mass. Moreover, assuming that all the Yukawa couplings are perturbative ($h^{ij}, y_{1,2} < 4\pi$), the neutrino sector of this theory may be probed by lattice calculations of $C_g$ or a more accurate measurement of the top mass. Furthermore, the parameter space of this concrete model can already be searched by gravitational waves from a confinement-induced phase transition, where future gravitational wave experiments at frequencies above the THz ballpark can further improve this search.

Finally, for this model example, all the dimensionful fundamental parameters are approximately in the order of $f \approx 1 \times 10^{18}$ GeV: for example, $6m_1 = 3m_{2,3} \approx M_i/2 = m_H = m_{H_\nu} = f$. Thus, a future perspective of this model may be to extend it to dynamically generate the vectorlike masses of the hyperfermions and the right-handed neutrino masses. In such a model, we propose an SU(6) completion of the elementary scalar sector similar to the SU(4) completion in the SU(4)/Sp(4) partially composite Higgs model presented in Ref. [15]. However, we have left a detailed study of this extension for future work.

## ACKNOWLEDGMENTS

I would like to thank G. Cacciapaglia and M. T. Frandsen for discussions. The idea of this work was developed at Fermilab operated by Fermi Research Alliance, LLC under Contract No. DE-AC02-07CH11359 with the United States Department of Energy. I would like to thank Fermilab and Caltech for hosting me during the development of this paper. Furthermore, this work was finished at CP$^3$-Origins at University of Southern Denmark and at Rudolf Peierls Centre for Theoretical Physics at University of Oxford. I acknowledge partial funding from The Council For Independent Research, Grant No. DFF 6108-00623. The CP3-Origins center is partially funded by the Danish National Research Foundation, Grant No. DNRF90. Finally, I acknowledge the funding from the Independent Research Fund Denmark, Grant No. DFF 1056-00027B.